\documentclass[showpacs,showkeys,prd,nofootinbib,amsmath,amssymb]{revtex4}

\usepackage{color}
\usepackage{graphicx}
\usepackage{physics}
\usepackage{bbm}

\usepackage{graphicx}
\usepackage{subfig}
\usepackage{amssymb}
\usepackage{amsmath}
\usepackage{bm}
\usepackage{color}
\usepackage{slashed}

\usepackage{lipsum}
\usepackage{booktabs}

\begin{document}

\def\half{\textstyle{1\over2}}
\def\third{\textstyle{1\over3}}
\def\quarter{\textstyle{1\over4}}
\newcommand{\ud}{\mathrm{d}}
\newcommand{\GN}{G_{\rm N}}

\title{Unveiling Mapping Structures of Spinor Duals}

\author{Cavalcanti, R. T.}
\email{rogerio.cavalcanti@unesp.br}
\affiliation{Departamento de F\'{\i}sica e Qu\'{\i}mica, \\ Universidade Estadual Paulista, UNESP,\\ Guaratinguet\'{a}, SP, Brazil.}

\author{Hoff da Silva,J. M.}
\email{julio.hoff@unesp.br}
\affiliation{Departamento de F\'{\i}sica e Qu\'{\i}mica, \\ Universidade Estadual Paulista, UNESP,\\ Guaratinguet\'{a}, SP, Brazil.}

\begin{abstract}
Following the program of investigation of alternative spinor duals potentially applicable to fermions beyond the standard model, we demonstrate explicitly the existence of several well defined spinor duals. Going further we define a mapping structure among them and the conditions under which sets of such dual maps do form a group. We also study the covariance of bilinear quantities constructed with the several possible duals, the invariant eigenspaces of those group elements and its connections with spinors classification, as well as dual maps defined as elements of group algebras. 
\end{abstract}



%\preprint{IPPP/04/80}
%\preprint{DCPT/04/160}

\maketitle

\section{Introduction} \label{sec:intro}

Spinors play a fundamental role in high energy physics since the very first theoretical development of the spin phenomena due to Pauli \cite{pauli}. The physics based on the spinor concept has become widely known after the outstanding work of Dirac in finding the dynamical equation for the electron \cite{dirac}. The mathematically formal concept of spinor, concerning its relation with the Minkowski space physics, on the other hand, may be traced back to the work of Cartan \cite{cartan}. The theory of spinors went through a huge development since then and both the mathematical and physical aspects acquired great relevance in the description of fermionic particles. 
One of the most important peculiarities of spinors, regarding the description of fermions, is that a given fermion cannot be detected alone \cite{weinberg}. The spinorial counterpart of such an assertion is that spinors, as mathematical objects carrying special representations of $SL(2,\mathbb{C})$, are indeed sensible to the double connectedness of Poincar\`e group, and, as such, shall furnish physically observable quantities only when composed with the spinor dual. Interestingly enough, the dual {\it a l\`a} Dirac was the only one widely studied until recently. The reason for that is quite obvious: there was little motivation for an alternative dual theory or further consideration, since the Dirac dual provides non null, real and Lorentz invariant orthonormality relations, engendering an appropriate quantum propagator, and furnishing the right locality structure for the usual fermionic quantum fields.   

It is somewhat recent the exploration of different fermions whose dynamics is dictated, up to the present knowledge, only by the Klein-Gordon equation, the so-called mass dimension one fermions \cite{ahluwa, Lee:2015sqj, Lee:2015jpa, Lee:2014opa, Villalobos:2015xca}. The main idea of such an approach is to pursue the construction of a quantum field theory candidate to dark matter \cite{jcaps,Ahluwalia:2009rh, Agarwal:2014oaa}. The theory underling these fermions starts by investigating special spinorial representations of $(0,1/2)\oplus(1/2,0)$ whose relative phases are fixed in such a way that the resulting spinors are eigenspinors of the charge conjugation operator, but not of the parity operator, as the usual Dirac spinors do. The former aspect is responsible for the neutrality of such spinors with respect to gauge charges, while the last one implies that they don't obey the Dirac dynamics. These spinors are the so-called Elko\footnote{Elko is an acronym of the German term ``Eigenspinoren des Ladungskonjugationsoperators'', which means ``eigenspinor of the charge conjugation operator''.}. Such spinors have been inducing a large amount of new research in recent years, from its formal \cite{daRocha:2011xb, Fabbri:2019vut, Cavalcanti:2014uta, daRocha:2011yr} and phenomenological aspects \cite{Rogerio:2019evl,Rogerio:2017gvr, Fabbri:2014foa, Fabbri:2010qv, Fabbri:2009aj, Alves:2014kta, Alves:2014qua} to a wide range of applications \cite{Villalobos:2018shc,Cavalcanti:2015nna, Pereira:2017bvq, Pereira:2016emd, S.:2014dja, Fabbri:2017xyk, Pereira:2018xyl,Pereira:2016muy}. The relevant point to our analysis here is that since the very early formulation it became clear the necessity of a more involved approach to reach the physically adequate Elko dual \cite{jcaps}. This necessity culminates with a theory for spinor dual \cite{Ahluwalia:2016rwl,epls, Rogerio:2017hwz, daSilvaa:2019kkt}, based on a judicious set of physical and formal requirements.    

In its first version, by requiring a non vanishing invariant norm, the Elko dual was given by $\bar{\bar{\psi}}=(\Xi\psi)^\dagger\eta$ (we reserve $\bar{\psi}$ to the Dirac dual, as usual). Invariance under Lorentz boosts and rotations sets $\eta=\gamma_0$, up to a irrelevant constant, while a real norm leads to \cite{jcaps,Ahluwalia:2016rwl} 
\begin{align}\label{xidef}
\Xi=m^{-1}\mathcal{G}(\phi)\gamma_\mu p^\mu,
\end{align} where, remarkably enough\footnote{See the explicit matrix form of $\mathcal{G}(\phi)$ in the Appendix \ref{matrices}.}, $\mathcal{G}(\phi)$ ($\phi$ being part of the spherical coordinates momentum parameterization) is invariant under the symmetries of $HOM(2)$ subgroup of the Lorentz group \cite{CYL, Lee:2015tcc}. This early version of the theory, so to speak, had non locality as a troublesome aspect. Non locality, however, is intrinsically related to the symmetry underling the theory. Hence, by requiring a Lorentz invariant spin sum it was demonstrated \cite{Ahluwalia:2016rwl} that the dual given by $\tilde{\psi}=(\Xi\psi)^\dagger\gamma_0 \mathcal{O}$ furnish a local Lorentz invariant theory\footnote{The $\mathcal{O}$ operator shall be denoted by $\mathcal{A}$ or $\mathcal{B}$ in the rest of the paper, in touch with original notation \cite{Ahluwalia:2016rwl}, as necessity appears.}.     

In this paper we are concerned to understand, following the judicious inspection of the Clifford algebra general definition of algebraic spinor duals introduced in Ref. \cite{daSilvaa:2019kkt}, links between different allowed spinor duals, the maps associating them, and unveil some hidden algebraic structures behind those maps. In Section II we study several duals appearing in somewhat recent papers and the maps connecting them.  In Section III we properly investigate the algebraic structures associated to the mappings previously defined, as group structures, invariant eigenspaces associated to the group elements, and the group algebras defined over the set of mappings connecting different duals. It includes the special case of invariant eigenspaces which shrinks the plethora of duals possibilities to the usual Dirac case and the covariance structure of the bilinear quantities constructed with several different duals. The Section IV is reserved to the conclusions. We leave for the Appendix the explicit matrix form of the operators (group elements) studied in Section III.

\section{Spinor duals and dual maps}\label{algebraic}

In order to elucidate the fundamental properties of spinor duals we shall first make use  the algebraic definition of spinors. As it is well known, algebraic spinors are properly defined as ideals of Clifford algebras. A Clifford algebra, on the other hand, is the algebraic structure upon which the Dirac theory is based on (see  Refs. \cite{Lounesto:2001zz} and \cite{Vaz:2016qyw} for a comprehensive introduction to Clifford algebras, spinors and applications in physics). Our aim in this section, however, is to briefly revisit the properties of the algebraic definition of spinors and its respective dual, as well as the construction of maps connecting different well defined duals. Since we are mainly interested in algebraic properties of spinor duals, 
along this paper we shall usually drop the helicity index and the space-time point dependence. %inasmuch as we are mainly interested in algebraic properties of spinor duals.

The Clifford algebra of the Minkowski space $\mathbb{M}\equiv \mathbb{R}^{1,3}$, denoted by $\mathcal{C}\ell_{1,3}$, is defined as  the associative unital algebra such that the Clifford application $\gamma:\mathbb{R}^{1,3} \to \mathcal{C}\ell_{1,3}$ is linear and satisfies\footnote{In this paper the Clifford basis shall be denoted, as usual, by $\gamma_\mu\equiv \gamma(\textbf{e}_\mu)$, where $\{\textbf{e}_\mu\}_{\mu=0}^3$ is the basis of the Minkowski space.}
\begin{align*}
\gamma (v)\gamma (u)+\gamma (u)\gamma (v)=2\eta({ v, u}), \qquad \forall\, { v,u} \in \mathbb{R}^{1,3},
\end{align*}
where $\eta$ is the Minkowski metric.
Algebraic spinors are minimal left ideals built upon primitive idempotents of the associated Clifford algebra. In fact, given the Clifford algebra $\mathcal{C}\ell_{1,3}$ and $f$ a primitive idempotent, the minimal left ideals are of the form $\mathcal{C}\ell_{1,3}f$. Analogously, minimal right ideals are also built upon primitive idempotents, having the form $f\mathcal{C}\ell_{1,3}$. Furthermore, a scalar can be obtained from $f\mathcal{C}\ell_{1,3}f$. It allows the definition of an inner product $\beta$ by  associating an arbitrary spinor $\psi$ (minimal left ideal) to its correspondent $\psi^\star$ (minimal right ideal), called the adjoint with respect to the inner product $\beta$, such that $\beta(\psi, \phi)=\psi^\star \phi \in \mathbb{R}$. 

Right ideals can be mapped into left ideals, and vice-versa, by involutions of the algebra. Idempotents, however, in general are not preserved by involutions. Namely, denoting an arbitrary involution by $\alpha$, follows $\alpha(\mathcal{C}\ell_{1,3} f)=\alpha(f)\,\alpha(\mathcal{C}\ell_{1,3})=\alpha(f)\,\mathcal{C}\ell_{1,3}$. Nevertheless, in general, $\alpha(f) \neq f$ and consequently $\alpha(f)\,\mathcal{C}\ell_{1,3} \neq f\,\mathcal{C}\ell_{1,3}$. Notwithstanding, there always exists an element $h\in \mathcal{C}\ell_{1,3}$ such that $\alpha(f)=h^{-1}f\,h$ and $\alpha(h)= h$ \cite{Vaz:2016qyw, Benn:1987fw}, allowing one to define $\psi^\star=h\,\alpha(\psi)=f\, h\,\alpha(\psi)$. Thus, an inner product is reached as 
\begin{equation}
\beta(\psi, \phi)=h\,\alpha(\psi)\phi=f\, h\,\alpha(\psi)\phi f \in f\mathcal{C}\ell_{1,3}f \simeq \mathbb{R},
\end{equation}
where $\alpha$ is called the adjoint involution of the inner product $\beta$. There are two natural involutions inside the structure of Clifford algebras, the reversion and Clifford conjugation \cite{Vaz:2016qyw}. These two involutions determine two non equivalent inner products, being any other inner product determined by an equivalent involution \cite{Benn:1987fw}. The algebra of interest for us is, in fact, the Dirac algebra, which is the complexification of $\mathbb{C}\ell_{1,3}$, denoted by
 $\mathbb{C}\otimes \mathbb{C}\ell_{1,3}$. Composing the complex conjugation with any other algebra involution generates a different adjoint involution, thus  a different inner product. The main adjoint involution in our case is the one equivalent to the hermitian conjugation on the algebra representation. For such it is sufficient that  $\alpha^\circledast(a)=h^{-1}a^\dagger h$ and $h^\dagger = h$, for any $a\in \mathbb{C}\otimes \mathcal{C}\ell_{1,3}$ and $h\in \mathbb{C}\otimes \mathcal{C}\ell_{1,3}$ \cite{Vaz:2016qyw, Benn:1987fw}. The adjoint (dual) spinor $\psi^\star$ thus reads
\begin{equation}\label{algdual}
\psi^\star=h\alpha^\circledast(\psi)\rightarrow\psi^\dagger h=[h\psi]^\dagger,
\end{equation}
where $\circledast$ denotes the complex conjugation in $\mathbb{C}\otimes \mathbb{C}\ell_{1,3}$ and $\psi^\star\phi \in \mathbb{C}$. As we have discussed in Ref. \cite{daSilvaa:2019kkt},  by taking $h=\gamma^0 \Xi$ (with $\Xi$ defined in Eq. \eqref{xidef}) the algebraic dual definition coincides with the one proposed in Ref. \cite{Ahluwalia:2016rwl}. The constrain on $h$, given by  $h^\dagger = h$, translates into $\Xi^\dagger \gamma^0=\gamma^0 \Xi$, which is obeyed by $\Xi$. A general definition of dual can thus be achieved by replacing $\Xi$ by a general operator $\Delta$ such that
\begin{align}\label{fundrel}
\Delta^\dagger \gamma^0=\gamma^0 \Delta.
\end{align}
 This is the fundamental constrain on $\Delta$ and furnish an explicit characterization to $\Delta$ as \cite{daSilvaa:2019kkt}

\begin{equation}
\Delta=\left[\begin{array}{cccc}
a_{11} & a_{12} & a_{13} & a_{14} \\ 
a_{21} & a_{22} & a^*_{14} & a_{24} \\ 
a_{31} & a_{32} & a^*_{11} & a^*_{21} \\ 
a^*_{32} & a_{42} & a^*_{12} & a^*_{22}
\end{array}\right], \;\; \text{with}\;\; a_{13},a_{31},a_{24},a_{42} \in \mathbb{R}. 
\end{equation}
The above matrix can, then, be displayed as a block matrix of the form
\begin{equation}\label{seis}
\Delta=\left[\begin{array}{cccc}
A & B  \\ 
C & A^\dagger 
\end{array}\right],\; \text{with}\; B^\dagger=B \;\;\text{and}\;\;  C^\dagger=C. 
\end{equation}
Hence, the general dual reads 
\begin{align}\label{gdual}
\psi^\star=[\gamma^0\Delta \psi]^\dagger=[\Delta \psi]^\dagger\gamma^0= \psi^\dagger\Delta^\dagger\gamma^0=\psi^\dagger\gamma^0\Delta.
\end{align}

Instead of constructing a dual of the above form, we could, alternatively, from the dual proposed in \cite{Ahluwalia:2016rwl} to define a dual map that preserves all the generality of $\Delta$. We shall denote such map by $\Omega$. There are two mainly advantages of using this approach. The first one is that, unlike $\Delta$, we can choose the dual map to explicitly depend on the momentum $p^\mu$ of the particle described by the spinor, using the $\Xi$ dependence on $p^\mu$ parameters. The second one is that, as we are going to introduce in the next section, when defined in that way, some unexpected algebraic structures of the set of $\Omega$ maps are unveiled. We want to make clear that the momentum dependency on the dual map is a choice, we could depart from a different dual associated to $\Omega=\mathbb{I}$ and have no momentum dependency at all.   The relationship between $\Delta$ and $\Omega$ is found by defining an arbitrary spinor dual $\psi^\star$ such that, 
\begin{align}\label{dualmap}
\psi^\star=[\Omega\gamma^0\Xi \psi]^\dagger=\psi^\dagger\gamma^0\Xi\Omega.
\end{align}
By comparing the Eqs. \eqref{gdual} and \eqref{dualmap} we find
\begin{align}\label{omegadef}
\gamma^0\Delta=\Omega\gamma^0\Xi,
\end{align}
or equivalently,
\begin{align}\label{omegadelta}
\Delta=\gamma^0 \Omega\gamma^0 \Xi \qquad \text{and} \qquad \Omega=\gamma^0 \Delta \Xi\gamma^0.
\end{align}
From \eqref{fundrel} and \eqref{omegadelta} follows the fundamental constrain of the $\Omega$ map
\begin{align} \label{fomega}
%\Omega^\dagger&=\gamma^0 \Xi^\dagger\Delta^\dagger \gamma^0\\
%&= \Xi\gamma^0\gamma^0\Delta \\
%&= \Xi\Delta \\
%&= \Xi\gamma^0 \Omega\gamma^0 \Xi.
\Omega^\dagger&=\gamma^0 \Xi^\dagger\Delta^\dagger \gamma^0 = \Xi\gamma^0\gamma^0\Delta = \Xi\Delta 
= \Xi\gamma^0 \Omega\gamma^0 \Xi.
\end{align}
%thus
Now we can check the $\Omega$ maps associated to each of the duals previously found in recent papers, accordingly:
\begin{itemize}
\item A possible dual discussed in Ref. \cite{jcaps} is $\psi^\star=[\Xi\psi]^\dagger \gamma^0$,
where
\begin{align}
\Omega=\mathbb{I} \qquad \text{and} \qquad  \Delta= \Xi.
\end{align}

\item The Dirac standard dual is $\psi^\star=\psi^\dagger \gamma^0$,
with
\begin{align}
\Omega=\gamma^0 \Xi\gamma^0=\Xi^\dagger \qquad \text{and} \qquad  \Delta= \mathbb{I}.
\end{align}

\item In Ref. \cite{Rogerio:2017hwz} it was found an equivalent Dirac dual given by $\psi^\star=[M_{\pm}\psi]^\dagger \gamma^0$, thus
\begin{align}
%\Omega&=\gamma^0 M_\pm \Xi\gamma^0\\
%&=\gamma^0 M_\pm \gamma^0\Xi^\dagger\\
%&=M_\pm^T\Xi^\dagger\\ 
%&=M_\mp\Xi^\dagger \qquad \text{and} \qquad \Delta= M_\pm.
\Omega&=\gamma^0 M_\pm \Xi\gamma^0=\gamma^0 M_\pm \gamma^0\Xi^\dagger = M_\pm^T\Xi^\dagger 
=M_\mp\Xi^\dagger \qquad \text{and} \qquad \Delta= M_\pm.
\end{align}

\item In Ref. \cite{Ahluwalia:2016rwl} it was also introduced $\psi^\star=[\mathcal{A}\Xi\psi]^\dagger \gamma^0$ and $\psi^\star=[\mathcal{B}\Xi\psi]^\dagger \gamma^0$, following
\begin{align}
\Omega=\gamma^0 \mathcal{A} \gamma^0 =\mathcal{A}\qquad \text{with} \qquad  \Delta= \mathcal{A}\Xi.
\end{align}
and
\begin{align}
\Omega=\gamma^0 \mathcal{B} \gamma^0=\mathcal{B}\qquad \text{with} \qquad  \Delta= \mathcal{B}\Xi.
\end{align}
respectively.
\end{itemize}
All those $\Omega$ maps are naturally in agreement with the constrain given by Eq. \eqref{fomega}.  

\section{Dual Maps and algebraic structures}

Now we are able to investigate the algebraic structure associated to the $\Omega$ maps. The first one is the possibility of a set of $\Omega$ maps do form a group. In order to form a group, a subgroup of GL(4,$\mathbb{C}$) which we shall denote generically by $G_{\Omega}$, such set must, as well known, obey the associativity, include an unit, be invertible and obey the closeness property. The associativity is straightforwardly inherited from the matrix algebra. The unit corresponds to the first case discussed in the previous Section. The invertibility is guaranteed by showing that, for a giving invertible $\Omega$ obeying Eq. \eqref{fomega}, the identity $\left(\Omega^{-1}\right)^\dagger=\Xi\gamma^0 \Omega^{-1} \gamma^0\Xi$ holds. In fact, from  
$\Omega^\dagger=\Xi\gamma^0 \Omega \gamma^0\Xi$ we have $\left(\Omega^\dagger\right)^{-1}=\Xi\gamma^0 \Omega^{-1} \gamma^0\Xi$. Therefore the result follows from $\left(\Omega^{-1}\right)^\dagger=\left(\Omega^\dagger\right)^{-1}$. 

The closeness is the less straightforward property. It also imposes a constrain on the possible $G_\Omega$ candidates. In fact, giving $\Omega_1$ and $\Omega_2$, from Eq. \eqref{fomega} we must have
\begin{align}\label{comega1}
\left(\Omega_1\Omega_2\right)^\dagger&=\Xi\gamma^0 \Omega_1 \Omega_2 \gamma^0\Xi.
\end{align}
On the other hand,
\begin{align}\label{comega2}
\left(\Omega_1\Omega_2\right)^\dagger&=\Omega_2^\dagger\Omega_1^\dagger=\Xi\gamma^0 \Omega_2 \gamma^0\Xi  \Xi\gamma^0 \Omega_1 \gamma^0\Xi
=\Xi\gamma^0 \Omega_2 \Omega_1 \gamma^0\Xi.
\end{align}
By comparing Eqs. \eqref{comega1} and \eqref{comega2} we conclude that $\Omega_1 \Omega_2=\Omega_2 \Omega_1$, thus $G_\Omega$ must be an Abelian subgroup of GL(4,$\mathbb{C}$). From $\Omega_1 \Omega_2=\Omega_2 \Omega_1$ and Eq. \eqref{omegadelta} we find the correspondent constrain on $\Delta$,
\begin{align}
 \gamma^0 \Delta_1 \Xi\gamma^0\gamma^0 \Delta_2 \Xi\gamma^0&=\gamma^0 \Delta_2 \Xi\gamma^0\gamma^0 \Delta_1 \Xi\gamma^0 \nonumber\\
\Delta_1 \Xi \Delta_2 \Xi&= \Delta_2 \Xi \Delta_1 \Xi,
% \qquad \Delta_1\Xi\Delta_2&=\Delta_2\Xi\Delta_1
\end{align} which by means of $\Xi^2=\mathbb{I}$ leads to 
\begin{align}
\Delta_1\Xi\Delta_2&=\Delta_2\Xi\Delta_1.
\end{align}

We could not find a simple form for the most general $G_\Omega$. Nevertheless, the particular cases introduced in the next Section are fairly straightforward and useful when connected to the Lounesto classification, as we shall show soon. 

\subsection{$G_\Omega$ Groups}

Once established the conditions for $G_\Omega$ being a group, we can now explore some particular cases explicitly. Below are listed all the $\Omega$ maps we are going to use as group elements, all of them are defined from compositions of the $\Xi$ operator and do satisfy the Eq. \eqref{fomega}. Note that, despite the different definition, the $\mathcal{G}(\phi)$ matrix is the same used in Eq. \eqref{xidef} for defining $\Xi$. The matrix form of such elements as well as some useful properties of them are found in the Appendix \ref{matrices}.% Here is the list: 

%****************
\begin{itemize}
\item $\mathcal{G}(\phi)=\mathcal{G}\equiv\frac{m}{2E}\{\gamma^0,\Xi\}=\frac{m}{2E}\left( \gamma^0\Xi+\Xi\gamma^0\right)$

\item $\mathcal{F}(\theta,\phi)=\mathcal{F}\equiv\frac{m}{2p}[\gamma^0,\Xi]=\frac{m}{2p}\left( \gamma^0\Xi-\Xi\gamma^0\right)$

\item $\mathcal{F}(\theta,\phi)\mathcal{G}(\phi)=\mathcal{F}\mathcal{G}=\frac{m^2}{4Ep}[\Xi^\dagger,\Xi]$

\item $\Xi^\dagger(p^\mu)=\Xi^\dagger=\gamma^0\Xi\gamma^0$

\item $\mathcal{G}\Xi^\dagger(p^\mu)=\frac{m}{2E}\left( \Xi^\dagger\Xi+\mathbb{I}\right)\gamma^0$

\item $\mathcal{H}(p^\mu)\equiv m^2\Xi\Xi^\dagger$

\item $\mathcal{H}^{-1}(p^\mu)=m^{-2}\Xi^\dagger\Xi=m^{-4}\gamma^0\mathcal{H}\gamma^0$

\end{itemize}
%*********************************

Three group structures are found by observing the above properties, two of them are straightforward given by $G_\mathcal{F}\equiv\{\mathbb{I},\mathcal{G},\mathcal{F},\mathcal{FG}\}$ and $G_{\Xi^\dagger}\equiv \{\mathbb{I},\mathcal{G},\Xi^\dagger,\mathcal{G}\Xi^\dagger\}$, whose Cayley tables are given below. 
Those groups are isomorphic to the classical Klein four group\footnote{For a basic reference on the subject, see \cite{klein}.} $K_4$. In spite of being isomorphic, $G_\mathcal{F}$ and $G_{\Xi}$ are topologically inequivalent. It comes from the fact that the $G_\mathcal{F}$ parameters are all compact.
\begin{table}[ht]
\centering
\begin{tabular}{c||cccc}
\hline 
\hline 
$G_\mathcal{F}$ & $\mathbb{I}$ & $\mathcal{G}$ & $\mathcal{F}$ & $\mathcal{F}\mathcal{G}$ \\ 
\hline 
\hline 
$\mathbb{I}$ & $\mathbb{I}$ & $\mathcal{G}$ & $\mathcal{F}$ & $\mathcal{F}\mathcal{G}$ \\ 
%\hline 
$\mathcal{G}$ & $\mathcal{G}$ & $\mathbb{I}$ & $\mathcal{F}\mathcal{G}$ & $\mathcal{F}$ \\ 
%\hline 
$\mathcal{F}$ & $\mathcal{F}$ & $\mathcal{F}\mathcal{G}$ & $\mathbb{I}$ & $\mathcal{G}$ \\ 
%\hline 
$\mathcal{F}\mathcal{G}$ & $\mathcal{F}\mathcal{G}$ & $\mathcal{F}$ & $\mathcal{G}$ & $\mathbb{I}$ \\ 
\hline
\hline  
\end{tabular} 
\qquad and \qquad %
\begin{tabular}{c||cccc}
\hline 
\hline 
$G_{\Xi^\dagger}$ & $\mathbb{I}$ & $\mathcal{G}$ & $\Xi^\dagger$ & $\Xi^\dagger\mathcal{G}$ \\ 
\hline 
\hline 
$\mathbb{I}$ & $\mathbb{I}$ & $\mathcal{G}$ & $\Xi^\dagger$ & $\Xi^\dagger\mathcal{G}$ \\ 
%\hline 
$\mathcal{G}$ & $\mathcal{G}$ & $\mathbb{I}$ & $\Xi^\dagger\mathcal{G}$ & $\Xi^\dagger$ \\ 
%\hline 
$\Xi^\dagger$ & $\Xi^\dagger$ & $\Xi^\dagger\mathcal{G}$ & $\mathbb{I}$ & $\mathcal{G}$ \\ 
%\hline 
$\Xi^\dagger\mathcal{G}$ & $\Xi^\dagger\mathcal{G}$ & $\Xi^\dagger$ & $\mathcal{G}$ & $\mathbb{I}$ \\ 
\hline
\hline  
\end{tabular} 
\caption{Cayley tables for $G_\mathcal{F}\equiv\{\mathbb{I},\mathcal{G},\mathcal{F},\mathcal{FG}\}$ (left panel) and $G_{\Xi^\dagger}\equiv \{\mathbb{I},\mathcal{G},\Xi^\dagger,\mathcal{G}\Xi^\dagger\}$ (right panel).}
\end{table}

%\bigskip\bigskip\bigskip\bigskip\bigskip\bigskip\bigskip\bigskip\bigskip

The remaining group, denoted by $G_\mathcal{H}$, is not of finite order as the previous ones. It is generated by $\{\mathbb{I},\mathcal{F},\mathcal{G},\mathcal{H},\mathcal{H}^{-1}\}$ and its Cayley table is given below, from where we can see $G_\mathcal{F}$ as a subgroup. We are going to use this fact later.

\begin{table}[ht]
\begin{tabular}{c||cccccccccc}
\hline 
\hline 
$G_\mathcal{H}$ & $\mathbb{I}$ & $\mathcal{G}$ & $\mathcal{F}$ &$\mathcal{FG}$& $\mathcal{H}$ &$\mathcal{H}^2$ & $\cdots$ & $\mathcal{H}^{-1}$  & $(\mathcal{H}^{-1})^2$ & $\cdots$ \\ 
\hline 
\hline 
$\mathbb{I}$ & $\mathbb{I}$ & $\mathcal{G}$ & $\mathcal{F}$ &$\mathcal{FG}$& $\mathcal{H}$ & $\mathcal{H}^2$ & $\cdots$ & $\mathcal{H}^{-1}$ & $(\mathcal{H}^{-1})^2$ & $\cdots$\\  
%\hline 
$\mathcal{G}$ & $\mathcal{G}$ & $\mathbb{I}$ & $\mathcal{FG}$ &$\mathcal{F}$& $\mathcal{GH}$ & $\mathcal{G}\mathcal{H}^2$ & $\cdots$ & $\mathcal{GH}^{-1}$ & $\mathcal{G}(\mathcal{H}^{-1})^2$ & $\cdots$\\   
%\hline 
$\mathcal{F}$ & $\mathcal{F}$ & $\mathcal{FG}$ & $\mathbb{I}$ &$\mathcal{G}$& $\mathcal{FH}$ & $\mathcal{F}\mathcal{H}^2$ & $\cdots$ & $\mathcal{FH}^{-1}$ & $\mathcal{F}(\mathcal{H}^{-1})^2$ & $\cdots$\\     
%\hline 
$\mathcal{FG}$ & $\mathcal{FG}$ & $\mathcal{F}$ & $\mathcal{G}$ &$\mathbb{I}$& $\mathcal{FGH}$ & $\mathcal{FG}\mathcal{H}^2$ & $\cdots$ & $\mathcal{FGH}^{-1}$ & $\mathcal{FG}(\mathcal{H}^{-1})^2$ & $\cdots$\\   
%\hline 
$\mathcal{H}$ & $\mathcal{H}$ & $\mathcal{GH}$ & $\mathcal{FH}$ &$\mathcal{FGH}$& $\mathcal{H}^2$ &$\mathcal{H}^3$ & $\cdots$ & $\mathbb{I}$ & $\mathcal{H}^{-1}$  &$\cdots$ \\ 
%\hline 
$\mathcal{H}^2$ & $\mathcal{H}^2$ & $\mathcal{G}\mathcal{H}^2$ & $\mathcal{F}\mathcal{H}^2$ &
$\mathcal{FGH}^2$& $\mathcal{H}^3$ &$\mathcal{H}^4$ & $\cdots$ & $\mathcal{H}$ & $\mathbb{I}$  &$\cdots$ \\ 
%\hline 
$\vdots$ & $\vdots$ & $\vdots$ &$\vdots$ & $\vdots$ &$\vdots$ & $\vdots$ & $\vdots$ & $\vdots$  &$\vdots$ &$\vdots$ \\   
%\hline 
$\mathcal{H}^{-1}$ & $\mathcal{H}^{-1}$ & $\mathcal{GH}^{-1}$ & $\mathcal{FH}^{-1}$ &
$\mathcal{FGH}^{-1}$& $\mathbb{I}$ &$\mathcal{H}^2$ & $\cdots$ & $(\mathcal{H}^{-1})^2$ &    $(\mathcal{H}^{-1})^3$  &$\cdots$ \\ 
%\hline 
$(\mathcal{H}^{-1})^2$ & $(\mathcal{H}^{-1})^2$ & $\mathcal{G}
(\mathcal{H}^{-1})^2$ & $\mathcal{F}(\mathcal{H}^{-1})^2$ &$\mathcal{FG}(\mathcal{H}^{-1})^2$& $\mathcal{H}^{-1}$ &$\mathbb{I}$ & $\cdots$ & $(\mathcal{H}^{-1})^3$ &    $(\mathcal{H}^{-1})^4$  &$\cdots$ 
\\ 
%\hline 
$\vdots$ & $\vdots$ & $\vdots$ &$\vdots$ & $\vdots$ &$\vdots$ & $\vdots$ & $\vdots$ & $\vdots$  &$\vdots$ &$\vdots$ \\  
\hline 
\hline 
\end{tabular} 
\caption{Cayley table for $G_\mathcal{H}\equiv\{\mathbb{I},\mathcal{F},\mathcal{G},\mathcal{H},\mathcal{H}^{-1},\cdots\}$.}
\end{table}

In principle, any element of these groups defines a different dual. However, as it could be expected, they are not all disconnected. Under certain circumstances discussed below the whole group defines a dual equivalent to the standard Dirac one, besides preserving the Lounesto Classes. At this point we face an important question regarding those new duals. Given that some of them are not equivalent to the Dirac dual nor any other previously discussed, may they have physical relevance? Specifically, a theory based of such dual is covariant? We shall study this fundamental aspect of the dual emerging from the above $\Omega$ groups in the next section.

\subsection{Investigating the Covariance}

After setting the several possibilities of duals and their group structure as well, we shall discuss the covariance of the resulting bilinear quantities. As the first usual step we recall that the spinors at hands belongs to a linear representation of the inhomogeneous Lorentz group and, therefore, must exist $S(\Lambda)$ invertible such that $\psi'(p')=S(\Lambda)\psi(p)$. To fix ideas let us restrict the analysis to transformations belonging to the orthochronous proper subgroup of the Lorentz group. Hereafter we shall denote $S(\Lambda)$ simply by $S$. Also, bearing in mind that the covariance of Dirac equation demand $S\gamma^\mu S^{-1}=\gamma^\nu \Lambda_\nu^{\;\;\mu}$ and $\gamma^0 S^{-1}=S^\dagger \gamma^0$, in order to achieve the right covariance for all bilinear quantities it is necessary and sufficient that the dual transform under $S$ as $\psi'^\star=\psi^\star S^{-1}$, as the usual Dirac case. Furthermore, the $\Xi$ operator is given by a specific sum of the type $\psi\bar{\psi}$ \cite{ahluwa,Ahluwalia:2016rwl},  and therefore $\Xi'=S\Xi S^{-1}$, leading to $(\Xi')^\dagger=(S^{-1})^\dagger\Xi^\dagger S^{\dagger}$.

Let us start by investigating the general $\Omega$ map dual $\psi^\star=[\Omega\gamma^0\Xi \psi]^\dagger$. According to the above discussion, under a symmetry transformation it must behave as $(\psi')^\star=\psi^\dagger \Xi^\dagger\gamma^0 \Omega^\dagger S^{-1}$, accordingly
\begin{align}
\psi'^\star&=\psi'^\dagger (\Xi')^\dagger\gamma^0 (\Omega')^\dagger\\
&=\psi^\dagger S^\dagger (S^{-1})^\dagger\Xi^\dagger S^{\dagger}\gamma^0 (\Omega')^\dagger \\
&=\psi^\dagger\Xi^\dagger \gamma^0 S^{-1} (\Omega')^\dagger 
\end{align}
Hence, in order to preserve covariance, $\Omega^\dagger$ must transform as $(\Omega')^\dagger =S\Omega^\dagger S^{-1}$, or equivalently, $\Omega'=(S^{-1})^\dagger\Omega S^{\dagger}$. Considering the group $G_{\Xi^\dagger}$, $\Omega=\Xi^\dagger$ automatically satisfies the correct transformation law. For $\Omega=\mathcal{G}$, from Eq. \eqref{xidef} it can be readly verified that $\mathcal{G}=m^{-1}\Xi\gamma^\mu p_\mu$. Noticing that $\mathcal{G}^\dagger=\mathcal{G}$ we have 
\begin{equation}
S^{-1}\mathcal{G}'S=m^{-1}S^{-1}\Xi'\gamma^\mu p'_\mu S=m^{-1}(S^{-1}\Xi'S)(S^{-1}\gamma^\mu p'_\mu S),
\end{equation} 
from which we see that $\mathcal{G}'=S\mathcal{G}S^{-1}$. As $\mathcal{G}\Xi^\dagger$ straightforwardly obey the same transformation law, we conclude that the group $G_{\Xi^\dagger}$ preserves the covariance.

The group $G_{\mathcal{F}}$ is a little more subtle. In fact, it is not possible to ensure the necessary transformation for any dual presenting $\mathcal{F}$ in its composition. The reason is the following. By using Eq. \eqref{xidef} and the matricial form for $\mathcal{G}$ and $\mathcal{F}$ (see the Appendix) a bit of algebra leads to  
\begin{equation}
\mathcal{F}=\frac{E}{mp}\Xi \gamma^\mu p_\mu-\frac{m}{p}\Xi\gamma^0.
\end{equation} 
Apart from the cumbersome coefficients (which shall not be taken as a necessary impediment to Lorentz covariance, but in this case are an element of trouble) the presence of $\gamma^0$ in the last term of the right hand side is problematic, preventing $\mathcal{F}$ to recast $\mathcal{F}'$ as $S\mathcal{F}S^{-1}$. Recall that the $S$ transformation is not unitary for Lorentz boosts (the boost sector of Lorentz transformations render it a non-compact group).

The results discussed so far may be partially applied to the cases of $G_{\mathcal{H}}$ but it needs some additional considerations. Firstly, the (infinite) elements of $G_{\mathcal{H}}$ containing $\mathcal{F}$ do not transform suitable and we cannot see their relevance yet, apart from mathematical aspects. The elements comprising $\mathcal{H}$ or  $\mathcal{H}^{-1}$ are quite interesting though. Notice that
\begin{align}
\mathcal{H}'&=m^2\Xi'(\Xi')^\dagger=m^{2}S\Xi S^{-1}(S^{-1})^\dagger \Xi^\dagger S^\dagger\\
(\mathcal{H}')^{-1}&=m^{-2}(\Xi')^\dagger\Xi'=m^{-2}(S^{-1})^\dagger \Xi^\dagger S^\dagger S\Xi S^{-1}.
\end{align}
In both cases the correct transformation is ensured if $S$ is unitary. Again, it is straightforwardly inherited by any power of $\mathcal{H}$ and $\mathcal{H}^{-1}$. As a result, the covariance of duals constructed with these elements requires unitarity of the spinorial transformation as the unique necessary and sufficient condition (recall that for $\mathcal{F}$ elements, there are additional inconvenience coming from the coefficients). As mentioned, this is not fulfilled in the scope of classical fields, but the case deserves special attention as it could be implemented in the framework of representations of the little group in the Hilbert space. Within this case, as it is well known, unitary and finite dimensional representations may certainly be found. The investigation of how to frame the different duals here studied requires further exploration, nevertheless the particular cases involving $\mathcal{H}$ or  $\mathcal{H}^{-1}$ could appear as physically relevant. Being more precise (remembering that we are concerned with orthochronous proper transformations), it is possible to achieve Lorentz transformations, say $L(p)$, connecting $k^\mu$ to $p^\mu$ by $p^\mu=L(p)^\mu_\nu k^\nu$ such that $p^2=k^2=m^2$. Moving forward, induced representations may be reached by means of special $W(\Lambda,p)$ elements given by $W(\Lambda,p)=L^{-1}(\Lambda p)\Lambda L(p)$ whose action on $k^\mu$ preserves it. The $W$ elements do form the little group. A set of spin one-half quantum states, $\{\Psi_{k,\pm 1/2}\}$, upon which the action of unitary and finite dimensional $U(W(\Lambda,p))$ engenders a genuine representation may well be defined. The relevance of duals built with $\mathcal{H}$ or $\mathcal{H}^{-1}$ would then be manifest in a mapping connecting the $\Psi_{k,\pm 1/2}$ set with its corresponding adjoint. This analysis is under in progress currently.  

Let us finish the section by taking a look at the subgroups of $G_\mathcal{H}$.  The first one, denoted by $G_{\mathcal{G}_0}$, is composed by $\{\mathbb{I},\mathcal{G}\}$. This subgroup, along with $G_{\Xi^\dagger}$ are in a Lorentz covariant sector, thus physically and mathematically
relevant. The second one, given by $G_{\mathcal{F}_0}=\{\mathbb{I},\mathcal{F}\}$ does not preserve Dirac nor unitary symmetry. The last one, $G_{\mathcal{H}_0}=\{\mathbb{I},\mathcal{H},\mathcal{H}^{-1},\mathcal{H}^2,\mathcal{H}^{-2},\ldots\}$, is associated to unitary symmetry and its structure opens new possibilities to be further explored. The hole group $G_\mathcal{H}$ can be decomposed in those subgroups, according to the associated symmetry\footnote{Such decomposition is formally well defined, since $G_{\mathcal{F}_0},G_{\mathcal{G}_0}$ and $G_{\mathcal{H}_0}$ are normal subgroups of $G_\mathcal{H}$ and $G_{\mathcal{F}_0}\cap G_{\mathcal{G}_0}\cap G_{\mathcal{H}_0}=\mathbb{I}$  \cite{fuchs1970infinite}}
\begin{align}
  G_\mathcal{H}=G_{\mathcal{F}_0}\oplus G_{\mathcal{G}_0}\oplus G_{\mathcal{H}_0}.
 \end{align}  

\subsection{$G_\Omega$ Groups and Invariant Eigenspaces}

Before to evince the structure of the related invariant eigenspaces, we shall say a few words about the so called Lounesto spinor classification \cite{Lounesto:2001zz} (see \cite{Cavalcanti:2014wia,HoffdaSilva:2017waf} for details). Roughly speaking, Lounesto shown that the bilinear covariants (composed with the usual Dirac dual) respecting the Fierz-Pauli-Kofink may serve to classify spinors. The idea is, via the inversion theorem \cite{inversion}, to use the values of the bilinear covariants to categorize spinors. This result in six disjoint different types of spinors, namely\footnote{Here the symbol $\gamma^{0123}$ stands for $\gamma^0\gamma^1\gamma^2\gamma^3$.}:
\begin{itemize}
\item Type (1) |   $\bar{\psi}\psi\neq 0$ and $\bar{\psi}\gamma^{0123}\psi\neq 0$;

\item Type (2) |  $\bar{\psi}\psi\neq 0$ and $\bar{\psi}\gamma^{0123}\psi=0$;

\item Type (3) | $\bar{\psi}\psi=0$ and $\bar{\psi}\gamma^{0123}\psi\neq 0$.
\end{itemize}
Elements of the above classes are called regular spinors, for which all the other bilinear covariants are non null. The remaining classes are called singular spinors. For them hold  $\bar{\psi}\psi= 0$ and $\bar{\psi}\gamma^{0123}\psi= 0$, along with:
\begin{itemize}
\item Type (4) |   $i\bar{\psi}\gamma^{0123}\gamma^\mu \psi\neq 0$ and $i\bar{\psi}\gamma^\mu\gamma^\nu\psi \neq 0$;

\item Type (5) |  $i\bar{\psi}\gamma^{0123}\gamma^\mu \psi= 0$ and $i\bar{\psi}\gamma^\mu\gamma^\nu\psi \neq 0$;

\item Type (6) | $i\bar{\psi}\gamma^{0123}\gamma^\mu \psi\neq 0$ and $i\bar{\psi}\gamma^\mu\gamma^\nu\psi=0$.  
\end{itemize}

Even though the Lounesto classification is build up based upon the standard Dirac dual, it is instructive to check the classes associated to the invariant subspaces (eigenspaces) of the new dual presented here. From elementary linear algebra it is well known that the eigenspaces are invariant under the action of commuting operators. It means that for any spinor belonging to those subspaces, the dual defined by its respective operator will match, up to a scalar (eigenvalue), the Dirac dual. If the dual is defined by a commuting operator, on the other hand, it will be equivalent to the Dirac dual for another spinor of the eigenspace. Given an operator $\mathcal{K}$, its correspondent eigenspace reads
\begin{align}
E_\lambda(\mathcal{K})=\left\{v \in \mathbb{C}^4 \;\vert\; \mathcal{K}v=\lambda v\right\}.
\end{align}
The eigenspaces of the $G_{\mathcal{F}}$, $G_{\Xi^\dagger}$ and $G_{\mathcal{H}}$ elements have  some characteristics in common\footnote{In this section we are considering the $G_\Omega$ elements as defining a dual through $\Delta$, as $\Delta=\mathcal{G}$, for example.}. For example, all the associated eigenvalues are degenerate with associated eigenspaces of dimension 2. In addition, the eigenvalues of $\mathcal{G}$, $\mathcal{F}$ and $\Xi^\dagger$ are $\pm 1$ and the eigenvectors belong to a well defined Lounesto classification, as discussed below.

The $\mathcal{G}$ operator has eigenspaces given by
\begin{align}
E_{\pm 1}(\mathcal{G})=\text{Span}_{\mathbb{C}}\left\{
\begin{pmatrix}
 \mp i e^{-i \phi }\\
 0\\
 0\\
 1 
\end{pmatrix}
,
\begin{pmatrix}
 0\\
 \pm i e^{i \phi }\\
 1\\
 0 
\end{pmatrix}
\right\}.
\end{align} 
An interesting feature of the above eigenspace is that any of its elements is of Type-(5) according to the Lounesto classification. In fact, the above eigenvectors are also eigenvectors of the charge conjugation operator with eigenvalues $\pm e^{-i\phi}, \pm e^{ i\phi}$ respectively \cite{Cavalcanti:2014wia}. As $\mathcal{G}$ commutes with any operator belonging to the $G_\Omega$ group introduced here, it follows that the Lounesto classification (Type-(5)) is preserved by any operator of the previous section.

For the $\mathcal{F}$ operator one finds the eigenspaces
\begin{align}
E_{\pm 1}(\mathcal{F})=\text{Span}_{\mathbb{C}}\left\{
\begin{pmatrix}
 \pm i e^{-i \phi }\cos \theta\\
 \pm i \sin \theta\\
 0\\
 1 
\end{pmatrix}
,
\begin{pmatrix}
\mp i \sin \theta\\
 \pm i e^{i \phi }\cos \theta \\
 1\\
 0 
\end{pmatrix}
\right\}.
\end{align}
In such case the eigenvectors, as well as their combinations, are regular, also according the Lounesto classification. Again, it means that any operator commuting with $\mathcal{F}$ and obeying the Eq. \eqref{fomega} will define a dual that maps regular spinors into regular spinors with the standard Dirac dual.

The eigenvectors of $\mathcal{H}$ and $\Xi^\dagger$ are all of Type-(6). However, as the Lounesto classification is in general not preserved by linear combinations, the eigenspace mix different Lounesto classes.
\begin{align}
E_{(\text{E}\mp p)^2}(\mathcal{H})=\text{Span}_{\mathbb{C}}\left\{
\begin{pmatrix}
0\\
0\\
  e^{-i \phi }(\cot \theta \pm \csc \theta)\\
 1 
\end{pmatrix}
,
\begin{pmatrix}
  e^{-i \phi }(\cot \theta \mp \csc \theta)\\
 1\\
 0 \\
 0
\end{pmatrix}
\right\}, 
\end{align}

\begin{align}
E_{\pm 1}(\Xi^\dagger)=\text{Span}_{\mathbb{C}}\left\{
\begin{pmatrix}
0\\
0\\
\frac{e^{-i \phi } \left(p \sin \theta\mp i m\right)}{\text{E}-p \cos \theta }\\
 1 
\end{pmatrix}
,
\begin{pmatrix}
-\frac{e^{-i \phi } \left(i m \pm p \sin \theta\right)}{\text{E}+p \cos \theta }\\
 1 \\
 0\\
 0 
\end{pmatrix}
\right\} .
\end{align}

\subsection{$G_\Omega$ Group Algebras}

The linearity of the Eq. \eqref{fomega} guarantee that combinations of $\Omega$ maps are also $\Omega$ maps. It turns the $G_\Omega$ groups into a wider algebraic structure, namely a group algebra \cite{lang2004algebra}. The Eq. \eqref{fomega} also restricts the scalars of the linear combination to belong to $\mathbb{R}$. Thus, as a group algebra element, $\Omega$ is of the form
\begin{align}
\Omega=\sum_{\omega \;\in\; G_\Omega} a_\omega \omega, \quad a_\omega \in \mathbb{R}. %\quad \text{and} \quad \omega \in G_\Omega.
\end{align}
The group algebras of the particular $G_\Omega$ groups presented here are indicated below. Note that the determinant gives the conditions on the real coefficients for $\Omega$ being invertible:

\begin{itemize}

\item $G_\Omega=G_{\Xi^\dagger}$ 
\begin{align}
\Omega&=a\mathbb{I}+b\Xi^\dagger+c\mathcal{G}+d\Xi^\dagger\mathcal{G},\\
\Delta&=  b\mathbb{I}+a\Xi+d\mathcal{G}+c\Xi\mathcal{G},\\
\det[\Omega]&=(a+b-c-d) (a-b+c-d) (a-b-c+d) (a+b+c+d).
\end{align}

\item $G_\Omega=G_\mathcal{F}$ 
\begin{align}
\Omega&=a\mathbb{I}+b\mathcal{F}+c\mathcal{G}+d\mathcal{F}\mathcal{G},\\
\Delta&= (a\mathbb{I}-b\mathcal{F}+c\mathcal{G}-d\mathcal{F}\mathcal{G})\Xi,\\
\det[\Omega]&=(a+b-c-d) (a-b+c-d) (a-b-c+d) (a+b+c+d).
\end{align}

\item $G_\Omega=G_\mathcal{H}$
\begin{align*}
\Omega&=a\mathbb{I}+b\mathcal{F}+c\mathcal{G}+d\mathcal{FG} +h_1\mathcal{H}+h_2\mathcal{H}^2+\cdots+\mathcal{F}(f_1\mathcal{H}+f_2\mathcal{H}^2+\cdots)+\\
&+\mathcal{G}(g_1\mathcal{H}+g_2\mathcal{H}^2+\cdots)+\mathcal{FG}(fg_1\mathcal{H}+fg_2\mathcal{H}^2+\cdots)+h_{-1}\mathcal{H}^{-1}+h_{-2}\mathcal{H}^{-2}+\cdots ,\\
\Delta&= \Xi[a\mathbb{I}+b\mathcal{F}+c\mathcal{G}+d\mathcal{FG} +h_1\mathcal{H}+h_2\mathcal{H}^2+\cdots+\mathcal{F}(f_1\mathcal{H}+f_2\mathcal{H}^2+\cdots)+\\
&+\mathcal{G}(g_1\mathcal{H}+g_2\mathcal{H}^2+\cdots)+\mathcal{FG}(fg_1\mathcal{H}+fg_2\mathcal{H}^2+\cdots)+h_{-1}\mathcal{H}^{-1}+h_{-2}\mathcal{H}^{-2}+\cdots].
\end{align*}

\end{itemize}
The possibility of $\Omega$ maps as composing group algebras increases considerably the possibility of duals, even for the very restrict set of $\Omega$ maps explicitly introduced here. Such possibility deserves a carefully attention in future investigations.  

\section{Concluding remarks and outlook}

In this paper we have introduced several possibilities with a rich algebraic structure for spinor duals, evincing different possibilities coming from solid foundations rooted on Clifford algebra. Rather than trying to find general properties of the allowed spinor duals, we focused on some interesting particular cases and its underlying group structure. We also clarify the connection between those group elements, its invariant eigenspaces and the so called Lounesto classification. The group algebra structure of the dual maps were also introduced.

 We would like to finalize this work by remarking some research paths which can be pursued in order to link the duals here investigated with physical theories. First of all, we shall emphasize that it seems indeed necessary to speculate about different duals. Apart from the known case previously mentioned there is also other possibilities\footnote{We are currently investigating whether the dual presented in Ref. \cite{rod} indeed respect the constraint imposed by Eq. (\ref{seis}) and, in a positive case, how this can be framed into the structure built in Sec. III.} arising in the scope of mass dimension one fermions \cite{rod}. The fulcrum of these investigations is the search for quantum field theory fermionic candidates to dark matter. Therefore, mathematically well posed duals seems to be a good start.   

Two aspects, one algebraic and other physical, concerning the duals possibilities here found shall be further explored. On the one hand, it would be important to investigate whether the Fierz-Pauli-Kofink identities, computed with the covariant bilinears constructed with these duals, holds. For positive cases, a Lounesto-like classification would be in order, whilst, for negative cases, one could relate the specific cases contrasting them to the so called amorphous spinors (sections of Clifford bundle which does not obey the FPK identities) \cite{wal,ori}. On the other hand, the explicit appreciation of spin sums resulting from the pair spinor/dual with the different cases here shown may be a secure rote to explore physical consequences of fermionic theories constructed upon different duals. In fact, by investigating the spin sums one shall appreciate from the physical invariance of the theory to the locality structure of the field in question, as well as the right canonical mass dimension via the quantum propagator.       

\subsection*{Acknowledgements}

RCT thanks the UNESP-Guaratinguet\'a Post-Graduation program and CAPES for the financial support. JMHS thanks to CNPq (grant no. 303561/2018-1) for partial support.

\appendix \label{matrices}

\section{Matrix Form of the $G_\Omega$ Elements}\label{matrices}

We depict here the explicit matrix form of the terms used in Sec. III for convenience, as well as some useful identities.

%%%%%%%%%%%%%%%%%%%%%%%%%%%%%%%%%%%%%%%G
\begin{minipage}[center]{0.5\textwidth}
\begin{small}
\begin{align*}
 \mathcal{G}(\phi)=\left[\begin{array}{cccc}
0 & 0 & 0 & -ie^{-i\phi} \\ 
0 & 0 & ie^{i\phi} & 0 \\ 
0 & -ie^{-i\phi} & 0 & 0 \\ 
ie^{i\phi} & 0 & 0 & 0
\end{array}\right],
\end{align*}
\end{small}
\end{minipage} \begin{minipage}[center]{0.5\textwidth}
\begin{small}
\begin{tabular}{ccccccc}
$\mathcal{G}^2=\mathbb{I};$&$\qquad$ & $\mathcal{G}^\dagger=\mathcal{G}; $&$\qquad$& $\det [\mathcal{G}]=1;$\\
$ [\mathcal{G},\gamma^0]=0;$ &$\qquad$& $[\mathcal{G},\mathcal{F}]=0;$ &$\qquad$&$ [\mathcal{G},\Xi^\dagger]=0;$ \\
$ [\mathcal{G},\mathcal{H}]=0$; &$\qquad$& $[\mathcal{G},\mathcal{H}^{-1}]=0$. 
\end{tabular}
\end{small}
\end{minipage}

%%%%%%%%%%%%%%%%%%%%%%%%%%%%%%%%%%%%%F
\begin{minipage}[center]{0.5\textwidth}
\begin{small}
\begin{align*} 
\mathcal{F}(\theta,\phi)=\left[
\begin{array}{cccc}
 0 & 0 & -\sin \theta & e^{-i \phi } \cos \theta \\
 0 & 0 & e^{i \phi } \cos \theta & \sin \theta \\
 \sin \theta & -e^{-i \phi } \cos \theta & 0 & 0 \\
 -e^{i \phi } \cos \theta & -\sin \theta & 0 & 0 \\
\end{array}
\right],
\end{align*}
\end{small}
\end{minipage} 
\begin{minipage}[center]{0.5\textwidth}
\begin{small}
\begin{tabular}{ccccccc}
$\mathcal{F}^2=\mathbb{I};$&$\qquad$ & $\mathcal{F}^\dagger=\mathcal{F}; $&$\qquad$& $\det [\mathcal{F}]=1;$ \\
$ \{\mathcal{F},\gamma^0\}=0;$ &$\qquad$& $[\mathcal{G},\mathcal{F}]=0;$ &$\qquad$&$ \{\mathcal{F},\Xi^\dagger\}=0;$\\
$ [\mathcal{F},\mathcal{H}]=0$; &$\qquad$& $[\mathcal{F},\mathcal{H}^{-1}]=0$. 
\end{tabular} 
\end{small}
\end{minipage}

%%%%%%%%%%%%%%%%%%%%%%%%%%%%%%%%%%%%%FG
\begin{minipage}[center]{0.5\textwidth}
\begin{small}
\begin{align*} 
\mathcal{F}\mathcal{G}&=\left[
\begin{array}{cccc}
 -\cos \theta  & -e^{-i \phi } \sin \theta  & 0 & 0 \\
 -e^{i \phi } \sin \theta  & \cos \theta  & 0 & 0 \\
 0 & 0 & \cos \theta  & e^{-i \phi } \sin \theta  \\
 0 & 0 & e^{i \phi } \sin \theta  & -\cos \theta  \\
\end{array}
\right],
\end{align*}
\end{small}
\end{minipage} 
\begin{minipage}[center]{0.5\textwidth}
\begin{small}
\begin{tabular}{ccccccc}
$\left(\mathcal{FG}\right)^2=\mathbb{I};$&$\qquad$ & $\left(\mathcal{FG}\right)^\dagger=\mathcal{FG}; $&$\qquad$& $\det [\mathcal{FG}]=1;$\\
 $ \{\mathcal{FG},\gamma^0\}=0;$  &$\qquad$& 
$[\mathcal{FG},\mathcal{F}]=0;$ &$\qquad$&$ \{\mathcal{FG},\Xi^\dagger\}=0;$ \\
$ [\mathcal{FG},\mathcal{H}]=0$; &$\qquad$& $[\mathcal{FG},\mathcal{H}^{-1}]=0$. 
\end{tabular} 
\end{small}
\end{minipage}

%%%%%%%%%%%%%%%%%%%%%%%%%%%%%%%%%%%%%Xi
\begin{minipage}[center]{0.5\textwidth}
\begin{footnotesize}
\begin{align*}
\Xi^\dagger=-\frac{i}{m}\left[
\begin{array}{cccc}
 { p \sin \theta } & { e^{-i \phi } (\text{E}-p \cos \theta)} & 0 & 0 \\
 { -e^{i \phi } (\text{E}+p \cos \theta)} & -{ p \sin \theta} & 0 & 0 \\
 0 & 0 & -{ p \sin \theta} & { e^{-i \phi } (\text{E}+p \cos \theta)} \\
 0 & 0 & -{ e^{i \phi } (\text{E}-p \cos \theta)} & { p \sin \theta} \\
\end{array}
\right],
\end{align*}
\end{footnotesize}
\end{minipage} 
\begin{minipage}[center]{0.5\textwidth}
\begin{footnotesize}
\begin{tabular}{ccccccc}
$(\Xi^\dagger)^2=\mathbb{I};$&$\qquad$ & $[\mathcal{G},\Xi^\dagger]=0; $\\
 $\det [\Xi^\dagger]=1;$ &$\qquad$& $ \{\mathcal{F},\Xi^\dagger\}=0$.
\end{tabular}
\end{footnotesize}
\end{minipage}

%%%%%%%%%%%%%%%%%%%%%%%%%%%%%%%%%%%%%GXi
\begin{minipage}[center]{0.5\textwidth}
\begin{footnotesize}
\begin{align*}
 \mathcal{G}\Xi^\dagger=\frac{1}{m}\left[
\begin{array}{cccc}
 { e^{i \phi} p \sin \theta } & \text{E}-p \cos \theta & 0 & 0 \\
 {  \text{E}+p \cos \theta} & e^{-i \phi }{ p \sin \theta} & 0 & 0 \\
 0 & 0 & - e^{i \phi }{ p \sin \theta} & { \text{E}+p \cos \theta} \\
 0 & 0 &  \text{E}-p \cos \theta & -{ e^{-i \phi }}{ p \sin \theta} \\
\end{array}
\right],
\end{align*}
\end{footnotesize}
\end{minipage} 
\begin{minipage}[center]{0.5\textwidth}
\begin{footnotesize}
\begin{tabular}{ccccccc}
$(\mathcal{G}\Xi^\dagger)^2=\mathbb{I};$&$\qquad$ & $(\mathcal{G}\Xi^\dagger)^\dagger=\mathcal{G}\Xi; $\\
$\det [\mathcal{G}\Xi^\dagger]=1;$ &$q\quad$& 
$[\mathcal{G},\mathcal{G}\Xi^\dagger]=0$\\
$ \{\mathcal{F},\mathcal{G}\Xi^\dagger\}=0$.
\end{tabular} 
\end{footnotesize}
\end{minipage}

%%%%%%%%%%%%%%%%%%%%%%%%%%%%%%%%%%%%%H
\begin{minipage}[center]{0.5\textwidth}
\begin{footnotesize}
\begin{align*}
\mathcal{H}=\left[
\begin{array}{cccc}
 \text{E}^2+2 p \cos \theta \text{E}+p^2 & 2 e^{-i \phi } \text{E} p \sin \theta & 0 & 0 \\
 2 e^{i \phi } \text{E} p \sin \theta & \text{E}^2-2 p \cos \theta \text{E}+p^2 & 0 & 0 \\
 0 & 0 & \text{E}^2-2 p \cos \theta \text{E}+p^2 & -2 e^{-i \phi } \text{E} p \sin \theta \\
 0 & 0 & -2 e^{i \phi } \text{E} p \sin \theta & \text{E}^2+2 p \cos \theta \text{E}+p^2 \\
\end{array}
\right],
\end{align*}
\end{footnotesize}
\end{minipage} 
\begin{minipage}[center]{0.5\textwidth}
\begin{footnotesize}
\begin{tabular}{ccc}
$\mathcal{H}=\mathcal{H}^\dagger;$&$\qquad$ & $\det [\mathcal{H}]=m^8;$\\
$[\mathcal{G},\mathcal{H}]=0;$ &$\qquad$& $[\mathcal{F},\mathcal{H}]=0.$
\end{tabular} 
\end{footnotesize}
\end{minipage}

%%%%%%%%%%%%%%%%%%%%%%%%%%%%%%%%%%%%%H^-1
\begin{minipage}[center]{0.5\textwidth}
\begin{footnotesize}
\begin{align*}
\mathcal{H}^{-1}=\left[
\begin{array}{cccc}
 {\text{E}^2-2 p \cos \theta \text{E}+p^2} & -2 e^{-i \phi } \text{E} p \sin \theta & 0 & 0 \\
 -2 e^{i \phi } \text{E} p \sin \theta& \text{E}^2+2 p \cos \theta \text{E}+p^2 & 0 & 0 \\
 0 & 0 & \text{E}^2+2 p \cos \theta \text{E}+p^2 & 2 e^{-i \phi } \text{E} p \sin \theta \\
 0 & 0 & 2 e^{i \phi } \text{E} p \sin \theta & \text{E}^2-2 p \cos \theta \text{E}+p^2
\end{array}
\right],
\end{align*}
\end{footnotesize}
\end{minipage} 
\begin{minipage}[center]{0.5\textwidth}
\begin{footnotesize}
\begin{tabular}{ccc}
$(\mathcal{H}^{-1})^\dagger=\mathcal{H}^{-1};$&$\quad$ & $\det [\mathcal{H}^{-1}]=\frac{1}{m^{8}};$\\
$[\mathcal{G},\mathcal{H}^{-1}]=0;$ &$\quad$& $[\mathcal{F},\mathcal{H}^{-1}]=0.$
\end{tabular}
\end{footnotesize}
\end{minipage}

%%%%%%%%%%%%%%%%%%%GH
\begin{scriptsize}
\begin{align*}
\mathcal{G}\mathcal{H}&=\left[
\begin{array}{cccc}
 0 & 0 & 2 i \text{E} p \sin \theta  & -i e^{-i \phi } \left(\text{E}^2+2 p \text{E}\cos \theta  +p^2\right) \\
 0 & 0 & i e^{i \phi } \left(\text{E}^2-2 p \text{E}\cos \theta  +p^2\right) & -2 i \text{E} p \sin \theta  \\
 -2 i \text{E} p \sin \theta  & -i e^{-i \phi } \left(\text{E}^2-2 p \text{E}\cos \theta  +p^2\right) & 0 & 0 \\
 i e^{i \phi } \left(\text{E}^2+2 p \text{E}\cos \theta  +p^2\right) & 2 i \text{E} p \sin \theta & 0 & 0 \\
\end{array}
\right],
\end{align*}
\end{scriptsize}

%%%%%%%%%%%%%%%%%%%FH
\begin{scriptsize}
\begin{align*}
\mathcal{F}\mathcal{H}&=\left[
\begin{array}{cccc}
 0 & 0 & -i \left(\text{E}^2+p^2\right) \sin \theta  & i e^{-i \phi } \left[2 \text{E} p+\left(\text{E}^2+p^2\right) \cos \theta \right] \\
 0 & 0 & i e^{i \phi } \left[\left(\text{E}^2+p^2\right) \cos \theta -2 \text{E} p\right] & i \left(\text{E}^2+p^2\right) \sin \theta  \\
 i \left(\text{E}^2+p^2\right) \sin \theta  & -i e^{-i \phi } \left[\left(\text{E}^2+p^2\right) \cos \theta -2 \text{E} p\right] & 0 & 0 \\
 -i e^{i \phi } \left[2 \text{E} p+\left(\text{E}^2+p^2\right) \cos \theta \right] & -i \left(\text{E}^2+p^2\right) \sin \theta  & 0 & 0 \\
\end{array}
\right]. 
\end{align*}
\end{scriptsize}

\bibliography{dualmaps}{}

\end{document}